\documentclass[aip,reprint]{revtex4-1}
\usepackage{graphicx}
\usepackage{dcolumn}
\usepackage{bm}

\input{tcilatex}
\begin{document}

\title{Transmission and reflection of charge density waves in a quantum Hall edge controlled by a metal gate}
\author{Masahiro Matsuura}
\affiliation{Department of Physics, Tohoku University, Sendai 980-8578, Japan}
\author{Takaaki Mano}
\affiliation{National Institute for Materials Science, Tsukuba, Ibaraki 305-0047, Japan}
\author{Takeshi Noda}
\affiliation{National Institute for Materials Science, Tsukuba, Ibaraki 305-0047, Japan}
\author{Naokazu Shibata}
\affiliation{Department of Physics, Tohoku University, Sendai 980-8578, Japan}
\author{Masahiro Hotta}
\affiliation{Department of Physics, Tohoku University, Sendai 980-8578, Japan}
\author{Go Yusa}
\email{yusa@tohoku.ac.jp}
\affiliation{Department of Physics, Tohoku University, Sendai 980-8578, Japan}
\date{\today }

\begin{abstract}
Quantum energy teleportation (QET) is a proposed protocol
related to the quantum vacuum. The edge channels in a quantum Hall system
is well suited for the experimental verification of
QET. For this purpose, we examine a charge density wave excited\ and
detected by capacitively coupled front gate electrodes. We observe the
waveform of the charge density wave, which is proportional to the time
derivative of the applied square voltage wave. Further, we study the
transmission and reflection behaviors of the charge density wave by applying
a voltage to another front gate electrode to control the path of the edge
state. We show that the threshold voltages where the dominant direction is
switched in either transmission or reflection for dense and sparse waves are
different from the threshold voltage where the current stops flowing in an
equilibrium state.
\end{abstract}

\maketitle

The physics of the quantum vacuum and its fluctuations (zero-point
fluctuations) have attracted considerable attention in various fields of
modern physics \cite{vacuum}. Quantum energy teleportation (QET) is one
quantum-vacuum-related protocol \cite{hotta,hottaJPA}. By this protocol, the
local zero-point energy is extracted from a remote place by only sending
classical information, which does not carry energy but contains how to
extract energy from the local vacuum. In order to verify this quantum
protocol by experiment, a quantum Hall (QH) system has been theoretically
suggested to be the best suited physical system \cite{yusa,hotta2014}. The
QH states consist of two regions---bulk and edge. When a strong
perpendicular magnetic field is applied to the two-dimensional (2D)
electrons, the orbital degree of freedom of the electrons in the bulk region
is quantized and does not contribute to the transport. However, electrons in
the edge region can flow without backscattering, leading to a zero
longitudinal resistance \cite{yoshioka}. Using intriguing properties of edge channels and
the charge density waves propagating along them, pioneering experiments have
been performed \cite{allen,ashoori,feve,kamata,hashisaka}. To perform the
QET protocol, excitation and detection through capacitively coupled gate
electrodes are required \cite{yusa,hotta2014}. In most experiments, however,
the charge density wave in the edge channel has been measured through ohmic
contacts as a current flow \cite{kamata,hashisaka}. The study of detection
by a capacitively coupled contact is very limited \cite{ashoori}. In this
paper, we establish a detection scheme for measuring the charge density wave
through a capacitively coupled gate electrode as a key ingredient for
experimental verification of the QET protocol.

\begin{figure}[t]
%orginal EPS size 0 0 241 171
%\includegraphics[bb=18 32 225 152, clip,width=8.6cm]{Fig1.eps}
%\includegraphics[bb=18 30 208 138, clip,width=8.6cm]{Fig1.eps}
%\includegraphics[bb=0 0 219 124, clip,width=8.6cm]{Fig1.eps} \caption{}%
%\includegraphics[bb=0 0 219 123, clip,width=8.6cm]{Fig1.eps}
%\includegraphics[clip,width=8.6cm]{Fig1.eps}
\par
\begin{center}
\includegraphics{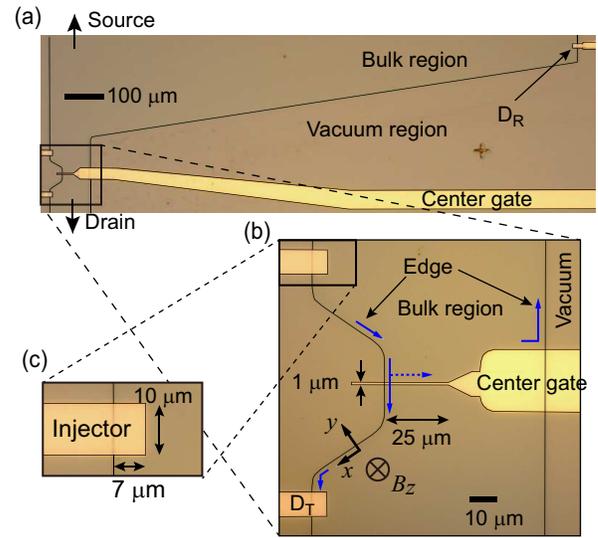}
\end{center}
\caption{(a) Optical microscope images of our device. (b) Close-up of the
center region. (c) Close-up of the injector, which is used to apply a
voltage and excite charge density waves in the edge channel.
 Detectors D$_{\text{T}}$, and D$_{\text{R}}$ are used to measure the density wave. 
By applying a negative center gate
voltage, $V_{\text{C}}$, the two-dimensional electrons beneath the center
gate can be depleted, and the paths of the wave packets can be switched
between detectors D$_{\text{T}}$, and D$_{\text{R}}$. The direction of the edge channel is indicated by
the blue arrows. The contours of the mesa and gate electrodes are outlined
for clarity.}
\label{fig:fig1}
\end{figure}

We used a GaAs/AlGaAs single heterostructure as a 2D system. The as-grown
electron density and mobility at $45$~mK are $2.3\times 10^{11}$~cm$^{-2}$
and $1.3\times 10^{6}$~cm$^{2}$/(Vs), respectively. Using photolithography
and electron-beam lithography, we fabricated the device shown in Fig.~1. The
center gate connected to a direct current (dc) voltage source can deplete
the electrons under the gate with the application of a negative center gate
voltage $V_{\text{C}}$. Thus, a change in $V_{\text{C}}$ can alter the
path of the edge channel from the injector to either of the two detectors, D$_{\text{T}}$
or D$_{\text{R}}$. We define transmission as the path along which a charge
density wave is transmitted through the center gate and reflection as the
path along which a charge density wave is reflected by the center gate. The
injector, D$_{\text{T}}$, and D$_{\text{R}}$ are front gate electrodes made
with $20$- and $80$-nm-thick Ti and Au. They are capacitively coupled with
the edge channels in an equivalent manner. Moreover, electron-beam
lithography was used to process them and the narrow region of the center
gate. A voltage signal from a function generator is transmitted to the
injector to excite a charge density wave, which propagates along the edge
channel in the directions indicated by blue arrows in Fig.~1(b). D$_{\text{T}%
}$ and D$_{\text{R}}$ are connected to an oscilloscope at room temperature
through a coaxial cable. The voltage signals measured by the $200$-MHz
bandwidth oscilloscope from D$_{\text{T}}$ and D$_{\text{R}}$ are $V_{\text{T%
}}$ and $V_{\text{R}}$, respectively. No filters or amplifiers are
introduced between the detectors and the oscilloscope. The experiment was
performed with a magnetic field $B_{z}=9.5$~T applied along the $z$ axis
perpendicular to the 2D plane. In this condition, the Landau-level filling
factor $\nu $ is unity. The typical base temperature is $\sim 45$~mK.

First, we tested the behavior of the center gate. We applied an alternating
voltage with an amplitude $V_{\text{SD}}=10$~mV oscillating at 13 Hz between
the source and the drain, i.e., two ohmic contacts. The current between
them, $I_{\text{SD}}$, was measured by the lock-in technique as a function
of $V_{\text{C}}$. By decreasing $V_{\text{C}}$, $I_{\text{SD}}$ decreases
at $V_{\text{C}}\sim -0.5$~V because the electrons under the center gate are
depleted, and the electron flow between the source and the drain is
hampered. Below $V_{\text{C}}<\sim -0.6$~V, $I_{\text{SD}}$ becomes $\sim 0$%
, indicating that the bulk regions connected to the source and drain are
separated from each other by the depleted region under the center gate.

\begin{figure}[t]
%orginal EPS size 0 0 241 171
%\includegraphics[bb=18 32 225 152, clip,width=8.6cm]{Fig1.eps}
%\includegraphics[bb=18 30 208 138, clip,width=8.6cm]{Fig1.eps}
%\includegraphics[bb=0 0 219 124, clip,width=8.6cm]{Fig1.eps} \caption{}%
%\includegraphics[bb=0 0 219 123, clip,width=8.6cm]{Fig1.eps}
%\includegraphics[clip,width=8.6cm]{Fig1.eps}
\par
\begin{center}
\includegraphics{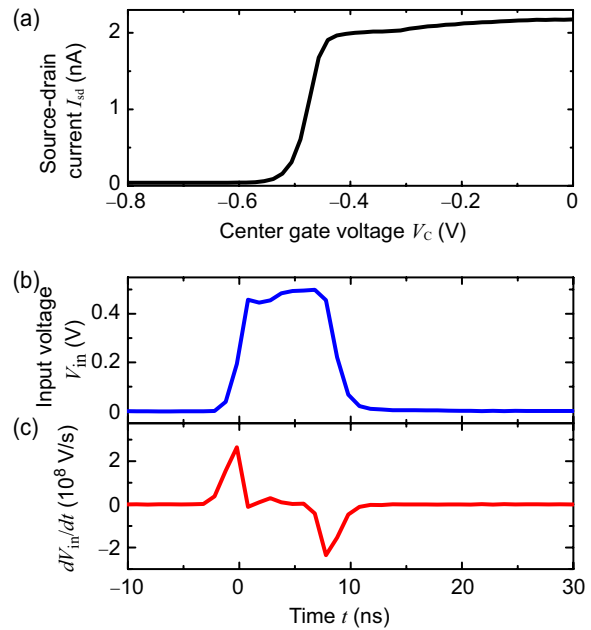}
\end{center}
\caption{(a) Source--drain current $I_{\text{sd}}$ dependence as a function
of the center gate voltage $V_{\text{C}}$ at $\protect\nu =1$, $B=9.5$ T,
and $T=40$ mK. (b) A typical waveform of the input voltage $V_{\text{in}}$
for exciting a charge density wave. The waveform was measured by connecting
a signal generator directly to an oscilloscope. (c) Time derivative $dV_{%
\text{in}}/dt $ of the waveform of Fig.~2(b). The time origin is arbitrary.}
\end{figure}

In order to excite a charge density wave, we used a square voltage wave. A
typical input voltage $V_{\text{in}}$ was measured by the oscilloscope by
direct connection to a signal generator. The typical rise and fall times of
the square wave are a few nanoseconds. The waveform has slight under- and
overshoots [Fig.~2(b)]. Thus, the time derivative $dV_{\text{in}}/dt$ of
this square wave has positive and negative peaks corresponding to the rising
and falling of the square wave, respectively, and has a small negative peak
right after the sharp positive peak at $t\sim 1$~ns in Fig.~$2$(c),
originating from the overshoot of the square wave. We discuss this waveform
of $V_{\text{in}}$ and $dV_{\text{in}}/dt$ later in relation to the
waveforms observed at D$_{\text{T}}$ and D$_{\text{R}}$.

We injected square waves with amplitudes of $0.5$~V and duration times of $5$%
~ns. With these square waves, a density wave is excited in the edge channel
and propagates along the edge, and we measure the waveforms at D$_{\text{T}}$
and D$_{\text{R}}$. We repeatedly performed single-shot measurements $6000$
times, and the averaged data are displayed in Fig.~3. When $V_{\text{C}}$ is
close to $\sim 0$~V, a voltage signal was only detected at D$_{\text{T}}$
[Fig.~3(a)], and no signal was detected at D$_{\text{R}}$ [Fig.~$3$(c)]. All
waves are transmitted through the center gate because the edge channel is
connected from the injector to D$_{\text{T}}$ but not to D$_{\text{R}}$. By
decreasing $V_{\text{C}}$, the amplitude of $V_{\text{T}}$ tends to become weak
[Fig.~$3$(b)], whereas that of $V_{\text{R}}$ becomes large [Fig.~3(d)],
suggesting that some part of the change density wave starts to propagate
along the edge formed by the center gate to reach the right-hand side of the
sample edge, which is connected to D$_{\text{R}}$. In other words, some part
of the charge density wave is reflected by the center gate. At $V_{\text{C}%
}=-0.8$~V, no signal was detected at D$_{\text{T}}$ because the center gate
is closed and all charge density waves are reflected by the center gate to D$%
_{\text{R}}$.

\begin{figure}[t]
%orginal EPS size 0 0 241 171
%\includegraphics[bb=18 32 225 152, clip,width=8.6cm]{Fig1.eps}
%\includegraphics[bb=18 30 208 138, clip,width=8.6cm]{Fig1.eps}
%\includegraphics[bb=0 0 219 124, clip,width=8.6cm]{Fig1.eps} \caption{}%
%\includegraphics[bb=0 0 219 123, clip,width=8.6cm]{Fig1.eps}
%\includegraphics[clip,width=8.6cm]{Fig1.eps}
\par
\begin{center}
\includegraphics{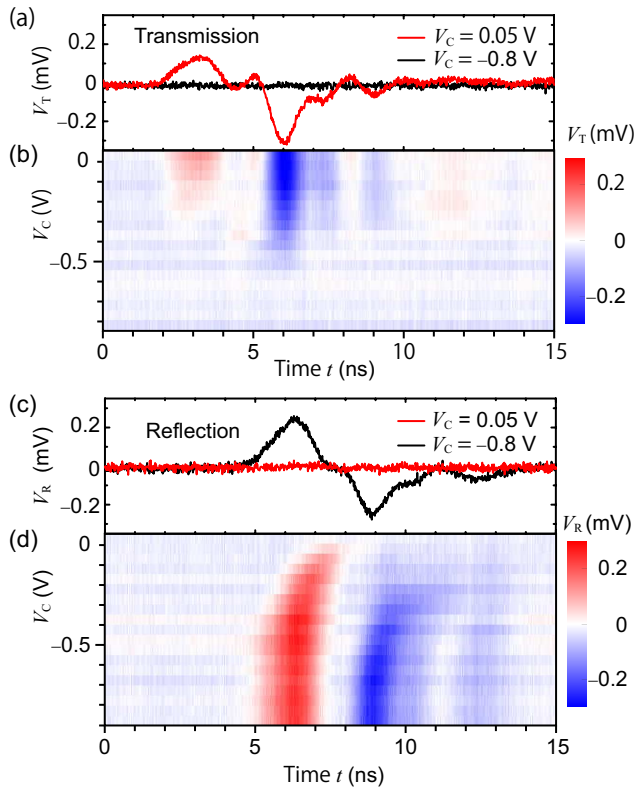}
\end{center}
\caption{(a) Waveforms of the voltage $V_{\text{T}}$ obtained at D$_{\text{T}%
}$ at $V_{c}=0.05$ V (black) and $-0.8$ V (red) as a function of the time $t$%
. (b) $V_{c}$ dependencies of the voltage waveforms obtained from D$_{\text{T%
}}$. (c) Waveforms of the voltage $V_{\text{R}}$ obtained at D$_{\text{R}}$
at $V_{\text{C}}=0.05$ V (black) and $-0.8$ V (red) as a function of the
time $t$. (d) Voltage waveforms obtained at D$_{\text{R}}$ as a function of $%
V_{\text{C}}$. All data were obtained at $\protect\nu =1$, $B=9.5$ T, and $%
\sim 40$ mK. The increment in $V_{\text{C}}$ for Figs.~2(b) and 2(d) is 50
mV.}
\end{figure}

The observed waveforms of $V_{\text{T}}$ and $V_{\text{R}}$ are not
analogous to the waveform of the applied voltage $V_{\text{in}}$ [Fig.~$2$%
(b)] but to that of $dV_{\text{in}}/dt$ [Fig.~$2$(c)]. When a square voltage
wave is applied to the injector, the electric field, i.e., the slope of the
local confinement potential $U(t)$, becomes small. Since the velocity $%
\upsilon (t)$ of the electrons in the edge channel is $\upsilon (t)=\frac{1}{%
|e|B}\frac{dU(t)}{dy}$ \cite{yoshioka}, $\upsilon (t)$ becomes temporarily
small during the rise time of the square pulse, which creates a local dense
region in the charge density wave. When the applied voltage becomes
constant, i.e., during the flat time region of the square pulse, $\upsilon
(t)$ is also constant over time, and the charge density is constant. Thus,
no charge density wave is created. During the fall time, a local sparse
region is created because $\upsilon (t)$ temporarily becomes large.
Therefore, the waveform of the charge density wave is proportional to $%
d\upsilon (t)/dt$. Note that since the QH state is incompressible, the
electron density in the bulk is constant. Thus, the change in the charge
density corresponds to the deformation of the edge \cite{yoshioka}. The
dense and sparse regions correspond to the convexity and concavity of the
boundary. Since D$_{\text{T}}$ and D$_{\text{R}}$ are capacitively coupled
to the edge channel, each waveform of $V_{\text{T}}$ and $V_{\text{R}}$ is
proportional to the charge density wave.

\begin{figure}[b]
%orginal EPS size 0 0 241 171
%\includegraphics[bb=18 32 225 152, clip,width=8.6cm]{Fig1.eps}
%\includegraphics[bb=18 30 208 138, clip,width=8.6cm]{Fig1.eps}
%\includegraphics[bb=0 0 219 124, clip,width=8.6cm]{Fig1.eps} \caption{}%
%\includegraphics[bb=0 0 219 123, clip,width=8.6cm]{Fig1.eps}
%\includegraphics[clip,width=8.6cm]{Fig1.eps}
\par
\begin{center}
\includegraphics{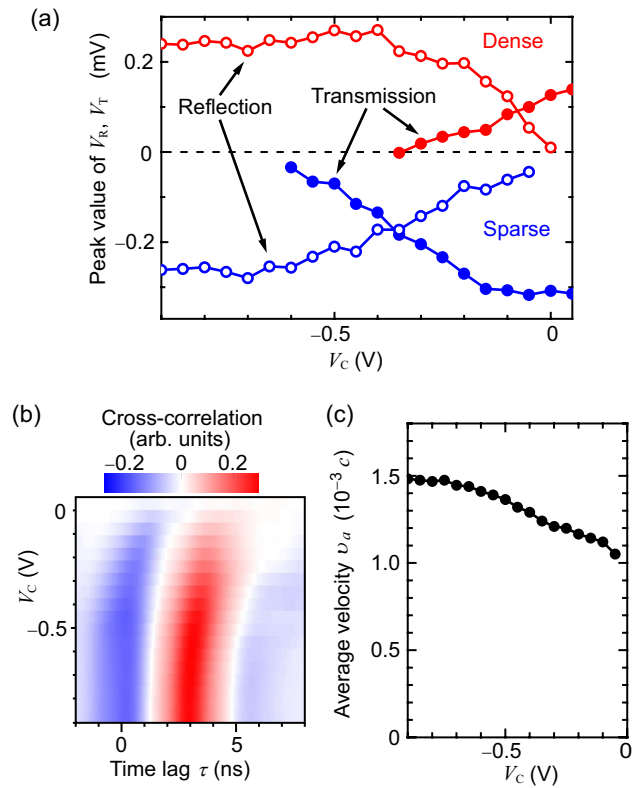}
\end{center}
\caption{(a) $V_{\text{C}}$ dependence of the amplitude of the local minima
(blue) and maxima (red) of $V_{\text{R}}$ (open circles) and $V_{\text{T}}$
(filled circles) in Figs.~3(b) and 3(d). (b) Cross-correlation of $V_{\text{T%
}}$ at $V_{\text{C}}=-0.05$~V and $V_{\text{R}}$ at each $V_{\text{C}} $ as
a function of the time lag $\protect\tau $. (b) Average velocity $\protect%
\upsilon _{a}$ of the charge density wave as a function of $V_{\text{C}}$ in
units of the speed of light in vacuum, $c\sim 3.00\times 10^{8}$ m/s.}
\end{figure}

The $V_{\text{C}}$ dependencies of the transmission and reflection of the
dense and sparse waves are notable. The peak values of $V_{\text{T}}$ and $%
V_{\text{R}}$ for dense and sparse waves are plotted as a function of $V_{%
\text{C}}$ in Fig.~4(a). For the dense wave (red), the transmission and
reflection cross each other at $V_{\text{C}}\sim -0.07$~V. In contrast, for
the sparse wave (blue), it requires a further negative voltage of $V_{\text{C%
}}\sim -0.36$~V for the transmission and reflection to cross each other.
Note that the value of $V_{\text{C}}$ at which $I_{\text{sd}}$ sharply
decreases is $V_{\text{C}}\sim -0.46$~V [Fig.~2(a)]. This may be because
some electrons can be excited to higher Landau levels when the dense region
is created by the rising part of the square wave. In contrast, the electrons
in the sparse region cannot be excited because there are no states. In other
words, the charge density waves are excited so strongly that the
electron--hole symmetry is broken. This is consistent with the fact that the
waveforms of the dense and sparse waves observed in reflection [the red line
in Fig.~3(c)] are more symmetric with respect to the 0-mV line than those
observed in transmission [the black line Fig. 3(a)]. This may be because all
of the electrons excited to the higher Landau levels can be easily relaxed
by the center gate, which functions as a scattering center, for $V_{\text{C}%
}=-0.8$~V, and they do not contribute to the reflection signal. In contrast,
for $V_{\text{C}}=0.05$~V, the influence of the center gate is small, and
the the dense wave is broader than the sparse wave in transmission.

The time difference $\delta t_{\text{R-T}}$ between the local maximum in $V_{%
\text{T}}$ at $V_{\text{C}}=-0.05$~V [Fig.~3(a)] and that in $V_{\text{R}}$
at $V_{\text{C}}=-0.8$~V [Fig.~3(c)] is $\sim 3$~ns and is equal to the
difference in the arrival times of the charge density waves at D$_{\text{T}}$
and D$_{\text{R}}$. Thus, $\delta t_{\text{R-T}}$ is $l_{\text{C}}/\upsilon
_{\text{C}}+(l_{\text{T}}-l_{\text{R}})/\upsilon $, where $l_{\text{T}}$, $l_{\text{R}}$, $%
l_{\text{C}}$, and $\upsilon _{\text{C}}$\ are the edge channel length
between the center gate to D$_{\text{T}}$, that between the center gate to D$%
_{\text{R}}$, that along the center gate, and the velocity of the edge
channel along the center gate, respectively. Here, $l_{\text{T}}=65$~$\mu $m, $%
l_{\text{R}}=1.3$~mm, and $l_{\text{C}}=65$~$\mu $m. Since $l_{\text{R}}>>l_{\text{T}},l_{\text{C}%
}$ and the order $O(\upsilon )\sim O(\upsilon _{\text{C}})$, $\delta t_{%
\text{R-T}}$ can be approximated by $l_{\text{R}}/\upsilon $, resulting in $%
O(\upsilon )\sim 4\times 10^{5}~$m/s ($\sim 1.3\times 10^{-3} $~$c$, where $%
c $ is the speed of light).

The local maximum and minimum in $V_{\text{T}}$ appear at $t\sim 3$ and $%
\sim 6$ ns, independent of $V_{\text{C}}$ [Fig.~3(b)], whereas those in $V_{%
\text{R}}$ shift as a function of $V_{\text{C}}$ for $V_{\text{C}}>\sim 0.4$%
~V and remain constant for $V_{\text{C}}<\sim 0.4$~V [Fig.~3(d)]. This may
be because the velocity changes when the charge density wave propagates
along the center gate electrode, as reported earlier \cite{kamata}. To
obtain $\delta t_{\text{R-T}}$ systematically, the cross-correlation of $V_{%
\text{T}}(t)$ at $V_{\text{C}}=-0.05$~V and $V_{\text{R}}(t)$ at each $V_{%
\text{C}}$, i.e., $\int V_{\text{T}}(t)V_{\text{R}}(\tau -t)dt$, is
calculated as a function of $V_{\text{C}}$. Here, $\tau $ is the time lag. A
strong positive correlation [the red region in Fig.~4(a)] appears near $\tau
\sim 3$~ns, which corresponds to $\delta t_{\text{R-T}}$. The average
velocity $\upsilon _{\text{avg}}=(l_{\text{C}}+l_{2})/\delta t_{\text{R-T}}$
between the center gate and D$_{\text{R}}$ is plotted as a function of $V_{%
\text{C}}$ in Fig.~4(c). By decreasing $V_{\text{C}}$, $\upsilon _{\text{avg}%
}$ gradually increases and becomes constant below $V_{\text{C}}<\sim -0.7$~V.

\begin{acknowledgments}
The authors are grateful for discussions with T. Fujisawa, K. Kobayashi, T.
Arakawa, J. N. Moore, R. Sch\"{u}tzhold, and W. G. Unruh. This work was supported by the
Asahi Glass Foundation;  MEXT/JSPS KAKENHI Grant Numbers JP 17H01037 and 16K05311(M.H.); and the  \textquotedblleft
Nanotechnology Support Project\textquotedblright\ of MEXT.
\end{acknowledgments}

% If in two-column mode, this environment will change to single-column format so that long equations can be displayed.
%\begin{widetext}
%$$\mbox{put long equation here}$$
%\end{widetext}

% Figures should be put into the text as floats.
% Use the graphics or graphicx packages (distributed with LaTeX2e).
% Put the label that you will use with \ref{} command in the braces of the \label{} command.
%
% \begin{figure}
% \end{figure}

% \begin{table}
% \end{tabular}
% \end{table}

%\begin{acknowledgments}
% Put your acknowledgments here.
%\end{acknowledgments}

\bibliographystyle{plain}
\bibliography{your-bib-file}

\begin{thebibliography}{99}
\bibitem{coment1}Although we only discuss the average velocity of charge density waves in this paper, one should note that the change of the velocity of the edge channel along the center gate, $\upsilon_{\text{C}}$,
 appears to be much bigger than reported earlier.
\end{thebibliography}

\end{document}